\documentclass[11pt]{article}

\usepackage{graphicx} 
\usepackage{amsmath}
\usepackage{amssymb}
\usepackage{bm}
\usepackage{braket}
\usepackage{amstext}
\usepackage{pdfpages}
\usepackage{amsxtra}
\usepackage{float}
\usepackage{upgreek}
\usepackage{times}
\usepackage{scicite}
\usepackage{caption}

\newcommand{\tg}{G}
\def\be{\begin{equation}}
\def\ee{\end{equation}}

\newenvironment{sciabstract}{%
\begin{quote} \bf}
{\end{quote}}

\newcounter{lastnote}
\newenvironment{scilastnote}{%
\setcounter{lastnote}{\value{enumiv}}%
\addtocounter{lastnote}{+1}%
\begin{list}%
{\arabic{lastnote}.}
{\setlength{\leftmargin}{.22in}}
{\setlength{\labelsep}{.5em}}}
{\end{list}}

\title{Critical Dynamics of Spontaneous Symmetry Breaking in a Homogeneous Bose gas}

\author
{Nir Navon$^{\ast \dag}$, Alexander L. Gaunt$^{\ast}$, Robert P. Smith, Zoran Hadzibabic\\ 
\\
\normalsize{Cavendish Laboratory, University of Cambridge,}\\
\normalsize{J. J. Thomson Avenue, Cambridge CB3 0HE, United Kingdom}\\
\\
\normalsize{$^{\ast}$These authors contributed equally to this work.}\\
\normalsize{$^\dag$To whom correspondence should be addressed; E-mail:  nn270@cam.ac.uk.}
}

\date{}

\begin{document} 


\baselineskip24pt


\maketitle 

\begin{sciabstract}
We explore the dynamics of spontaneous symmetry breaking in a homogeneous system by thermally quenching an atomic gas with short-range interactions through the Bose-Einstein phase transition. Using homodyne matter-wave interferometry to measure first-order correlation functions, we verify the central quantitative prediction of the Kibble-Zurek theory, namely the homogeneous-system power-law scaling of the coherence length with the quench rate. Moreover, we directly confirm its underlying hypothesis, the freezing of the correlation length near the transition due to critical slowing down. Our measurements agree with beyond mean-field theory, and support the previously unverified expectation that the dynamical critical exponent for this universality class, which includes the $\lambda$-transition of liquid  $^4$He, is $z=3/2$.

\end{sciabstract}


Continuous symmetry-breaking phase transitions are ubiquitous, from the cooling of the early universe to the $\lambda$-transition of superfluid helium. Near a second-order transition, critical long-range fluctuations are characterized by a diverging correlation length $\xi$ and details of the short-range physics are largely unimportant. Consequently, all systems can be classified into a small number of universality classes, according to their generic features such as symmetries, dimensionality and range of interactions~\cite{Kardar:2007}. Close to the critical point, many physical quantities exhibit power-law behavior governed by critical exponents characteristic of a universality class. Specifically, for a classical phase transition, $\xi \sim |(T-T_c)/T_c|^{-\nu}$, where $T_c$ is the critical temperature and $\nu$ is the (static) correlation-length critical exponent. Importantly, the corresponding relaxation time $\tau$, needed to establish a diverging $\xi$, also diverges: $\tau \sim \xi^z$, where $z$ is the dynamical critical exponent~\cite{Hohenberg:1977}.
An elegant framework for understanding the implications of this {\it critical slowing down} for the dynamics of symmetry breaking is provided by the Kibble-Zurek (KZ) theory \cite{Kibble:1976,Zurek:1985}.

\begin{figure} [b]
\centerline{\includegraphics[width=0.5\columnwidth]{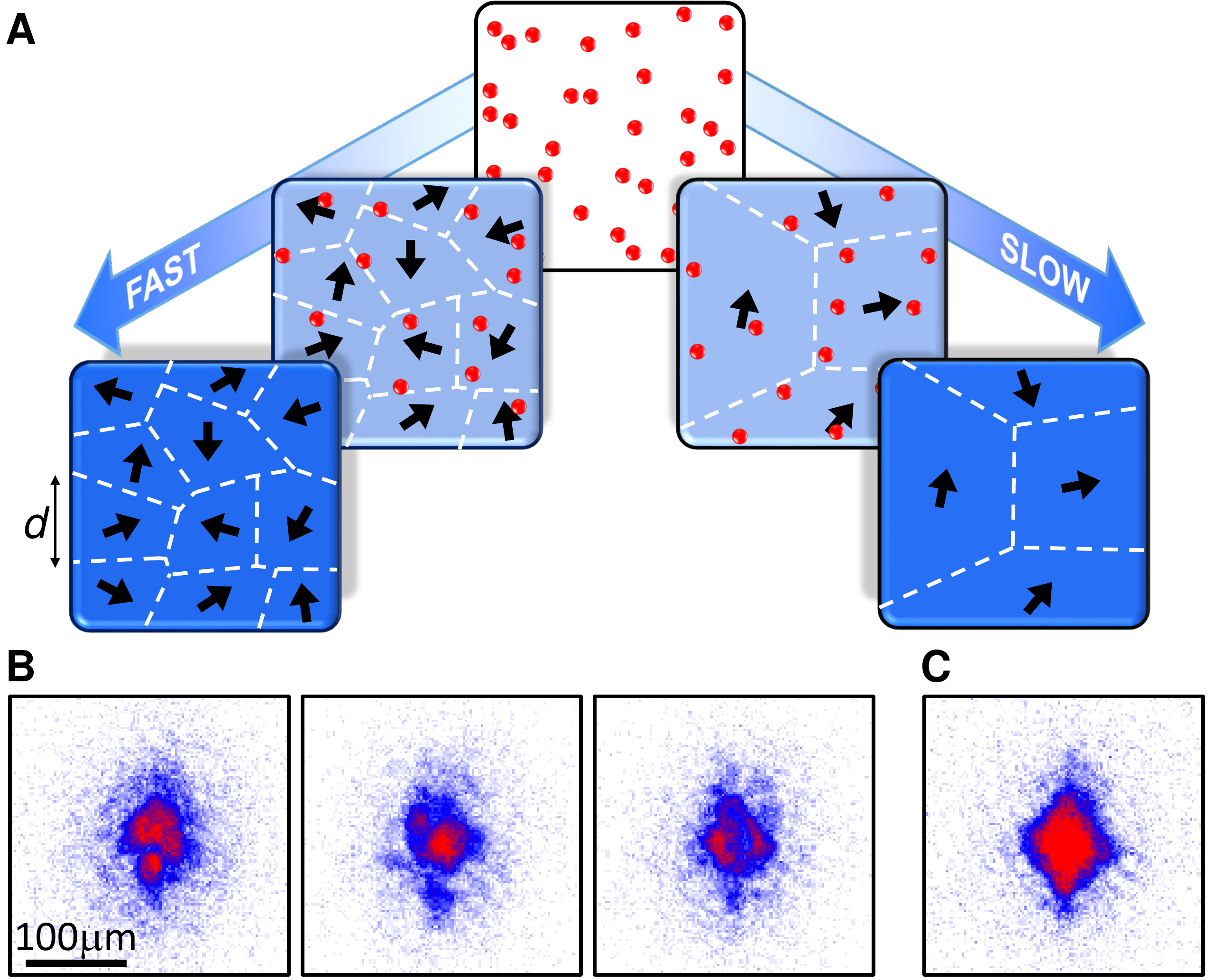}}
\caption{ {\bf Domain formation during spontaneous symmetry breaking in a homogeneous Bose gas}. (A) Red points depict thermal atoms and blue areas coherent domains, in which the $U(1)$ gauge symmetry is spontaneously broken. The arrows indicate the independently chosen condensate phase at different points in space, and dashed lines delineate domains over which the phase is approximately constant. The average size $d$ of the domains formed at the critical point depends on the cooling rate. Further cooling can increase the population of each domain before the domain boundaries evolve. (B) Phase inhomogeneities in a deeply degenerate gas are revealed in time-of-flight expansion as density inhomogeneities. Here the gas is cooled in 1~s from $T\approx 170$~nK, through $T_c \approx 70$~nK, to $\lesssim 10$ nK. Each realization of the experiment results in a different pattern, and averaging over many images results in a smooth featureless distribution. (C) Preparing a $T \lesssim 10$ nK gas more slowly (over 5~s) results in an essentially pure BEC with a spatially uniform phase.}
\label{fig:cartoon}
\end{figure}

Qualitatively, as $T$ is reduced towards $T_c$ at a finite rate, beyond some point in time the correlation length can no longer adiabatically follow its diverging equilibrium value. Consequently, at time $t=t_c$, the transition occurs without $\xi$ ever having spanned the whole system. This results in the formation of finite-sized domains that display independent choices of the symmetry-breaking order parameter, as illustrated in Fig.~\ref{fig:cartoon}A. (At the domain boundaries, rare long-lived topological defects can also form~\cite{DelCampo:2014}, their nature and density depending on the specific physical system.) 
Such domain formation was discussed in a cosmological context, and linked to relativistic causality, by Kibble~\cite{Kibble:1976}, while the connection to laboratory systems, critical slowing down and universality classes was made by Zurek~\cite{Zurek:1985}.

The main quantitative prediction of the KZ theory is that, under some generic assumptions~\cite{DelCampo:2014}, the average domain size $d$ follows a universal scaling law. The crucial KZ hypothesis is that in the non-adiabatic regime close to $t_c$ the correlations remain essentially frozen. Then, for a smooth temperature quench, the theory predicts
\begin{equation}
d = \lambda_0 \left( \frac{ \tau_Q}{\tau_0} \right)^b \, ,
\label{eq:KZ}
\end{equation}
with the KZ exponent
\begin{equation}
b = \frac{\nu}{1+ \nu z} \, ,
\label{eq:b}
\end{equation}
where $\tau_Q$ is the quench time defined so that close to the transition $T/T_c = 1+(t_c-t)/\tau_Q$, and $\lambda_0$  and $\tau_0$ are a system-specific microscopic length- and time-scale, respectively. 

Signatures of Kibble-Zurek physics have been observed in a wide range of systems, including liquid crystals~\cite{Chuang:1991}, liquid helium~\cite{Bauerle:1996,Ruutu:1996}, superconductors~\cite{Carmi:1999,Carmi:2000,Monaco:2002}, atomic Bose-Einstein condensates (BECs)~\cite{Sadler:2006,Weiler:2008,Chen:2011,Lamporesi:2013,Braun:2014,Corman:2014}, multiferroics~\cite{Chae:2012} and trapped ions \cite{Ulm:2013,Pyka:2013,Ejtemaee:2013}. 
However, despite this intense activity, a direct quantitative comparison with Eqs.~(\ref{eq:KZ}-\ref{eq:b}) has remained elusive; some common complications include system inhomogeneity, modified statistics of low-probability defects, and uncertainties over the nature of the transition being crossed (for a recent review see  \cite{DelCampo:2014}). In this work, we study the dynamics of spontaneous symmetry breaking in a homogeneous atomic Bose gas, which is in the same universality class as 3D superfluid $^4$He. For this class, mean-field (MF) theory predicts $\nu =1/2$ and $z=2$, giving $b=1/4$, while the beyond-MF dynamical critical theory, the so-called {\it F model}~\cite{Hohenberg:1977}, gives $\nu \approx 2/3$ and $z=3/2$, so $b\approx1/3$. We report the observation of the homogeneous-system KZ scaling law with the exponent $b$ in agreement with the beyond-MF theory. Moreover, using different quench protocols, we identify the regime of applicability of the KZ scaling law, and directly demonstrate the central role played by the freezing of the correlations near $t_c$.

We prepare a homogeneous Bose gas by loading $3\times 10^5$ $^{87}$Rb atoms into a cylindrical optical-box trap~\cite{Gaunt:2013} of length $L\approx 26\,\mu$m along the horizontal $x$-axis, and radius $R\approx 17\,\mu$m. Initially $T\approx 170$~nK, corresponding to $T/T_c \approx 2$. We then evaporatively cool the gas by lowering the trap depth, cross $T_c \approx 70$~nK with $2\times 10^5$ atoms, and have $10^5$ atoms at $T\lesssim 10$~nK ($T/T_c \lesssim 0.2$).
In our system $\lambda_0$ is expected to be set by the thermal wavelength at the critical point, $\lambda_c \approx 0.7\,\mu$m, and $\tau_0$ by the elastic scattering time $\tau_{\rm el}$~\cite{Anglin:1999, Weiler:2008}; for our parameters, a classical estimate gives $\tau_{\rm el} \approx 30$~ms. 

Qualitatively, random phase inhomogeneities in rapidly quenched clouds are revealed in time-of-flight (ToF) expansion as density inhomogeneities~\cite{Dettmer:2001,Chen:2011}, such as shown in Fig.~\ref{fig:cartoon}B (here the gas was cooled to $T \ll T_c$ in $1$~s). In our finite-sized box, we can also produce essentially pure and fully coherent (single-domain) BECs, by cooling the gas slowly (over $\gtrsim 5$~s). In ToF, such a BEC develops the characteristic diamond shape~\cite{Gotlibovych:2014} seen in Fig.~\ref{fig:cartoon}C.

\begin{figure*} [ht]
\includegraphics[width=\textwidth]{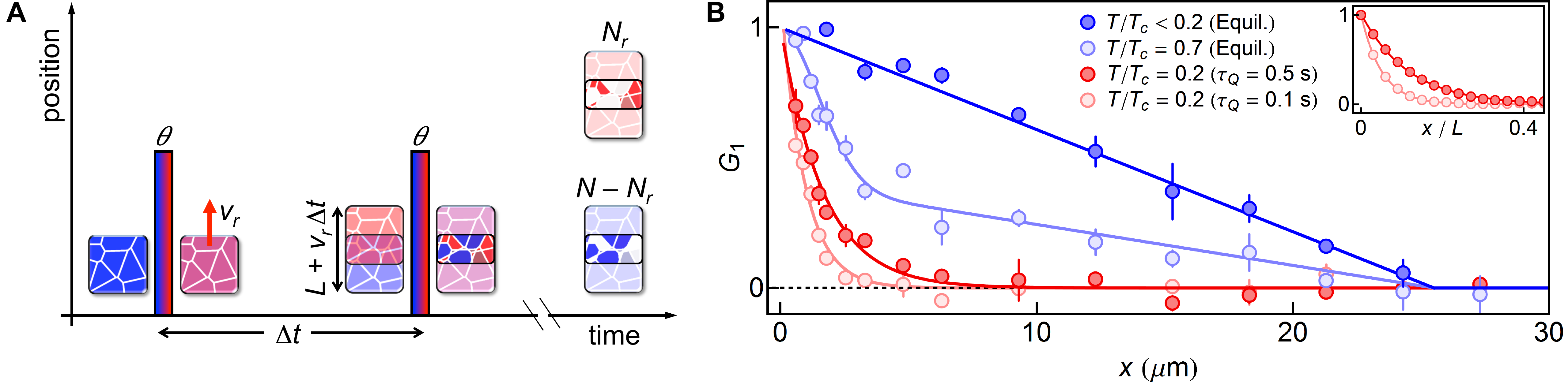}
\caption{ {\bf Two-point correlation functions in equilibrium and quenched gases}. (A) Homodyne interferometric scheme. The first Bragg-diffraction pulse ($\theta$) creates a superposition of a stationary cloud and its copy moving with a centre-of-mass velocity $v_r$. After a time $\Delta t$, a second pulse is applied. In the region where the two copies of the cloud displaced by $x=v_r\Delta t$ overlap, the final density of the diffracted atoms depends on the relative phase of the overlapping domains; $g_1(x)$ is deduced from the diffracted fraction $N_r/N$ (see text). (B) Correlation function $\tg_1 (x) = (1-x/L)g_1(x)$ measured in equilibrium (blue) and after a quench (red) for, respectively, two different $T/T_c$ values and two different quench times. Inset: 1D calculation of $G_1$ for a fragmented BEC containing $\mathcal{D} = 10$ (red) and $20$ (light red) domains of random sizes and phases. The solid lines correspond to $g_1=\exp(-x\mathcal{D}/L)$.}
\label{fig:g1}
\end{figure*}

To quantitatively study the coherence of our clouds we probe the first-order two-point correlation function
\begin{equation}
g_1({\bf r},{\bf r'}) \propto \langle \hat{\Psi}^\dag({\bf r}) \hat{\Psi}({\bf r'})\rangle \, ,
\label{eq:g1}
\end{equation}
where $\hat{\Psi}({\bf r})$ is the Bose field. Our method, outlined in  Fig.~\ref{fig:g1}A, is inspired by Ref.~\cite{Hagley:1999b}. We use a short (0.1 ms) Bragg-diffraction light pulse to create a small copy of the cloud (containing $\approx 5\%$ of the atoms) moving along the $x$-axis with recoil velocity $v_r \approx 3$~mm/s~\cite{Gotlibovych:2014}. A second identical pulse is applied a time $\Delta t$ later, when the two copies are shifted by $x = v_r \Delta t$, and for  $x< L$ still partially overlap. This results in interference of the two displaced copies of the cloud in the overlap region of length $L-x$. After the second Bragg pulse the fraction of diffracted atoms (for $x<L$) is \cite{SOM}
\begin{equation}
\frac{N_r}{N} = \frac{1}{2} \left[ 1+ \left(  1- \frac{x}{L}\right) g_1(x) \right] \sin^2 \theta \,,
\label{eq:Nr}
\end{equation}
where $g_1(x)\equiv \operatorname{Re}[g_1({\bf r},{\bf r}+x~{\bf \hat{x}})]$ is the correlation function corresponding to periodic boundary conditions and normalized so that $g_1(0)=1$, and $\theta$ is the area of each Bragg pulse (in our case $\theta \approx \pi/7$). Allowing the recoiling atoms to fully separate  from the main cloud (in 140~ms of ToF) and counting $N_r$ and $N$, we directly measure $\tg_1(x) \equiv (1-x/L) g_1(x)$, with a spatial resolution of $\approx 0.7\,\mu$m. Our resolution is limited by the duration of the Bragg pulses and the (inverse) recoil momentum; we experimentally assessed it by measuring $G_1$ in a thermal cloud with a thermal wavelength $<0.5\,\mu$m.

In Fig.~\ref{fig:g1}B we show examples of $\tg_1(x)$ functions measured in equilibrium (blue) and after a quench (red).
In an essentially pure equilibrium BEC (prepared slowly, as for Fig.~\ref{fig:cartoon}C), $g_1(x)=1$ and $G_1(x)$ is simply given by the triangular function $1- x/L$ (dark blue solid line). In equilibrium at $T/T_c \approx 0.7$, we see a fast initial decay of $\tg_1$, corresponding to the significant thermal fraction. However, importantly, the coherence still spans the whole system, with the slope of the long-ranged part of $\tg_1$ giving the condensed fraction (light blue line is a guide to the eye). By comparison, the $\tg_1$ functions for quenched clouds clearly have no equilibrium interpretation. Here $T/T_c \approx 0.2$, corresponding to a phase space density $>25$, and yet coherence extends over only a small fraction of $L$. These data are fitted well by $g_1 \propto \exp(-x/\ell)$ (red lines), which provides a simple and robust way to extract the coherence length.
This exponential form is further supported by a 1D calculation shown in the inset of Fig.~\ref{fig:g1}B. 
Here we generate a wavefunction with a fixed number of domains $\mathcal{D}$, randomly positioning the domain walls and assigning each domain a random phase. Averaging over many realizations, we obtain $g_1(x)$ that is fitted very well by an exponential with $\ell = L/\mathcal{D} =d$. (In our 3D experiments the total number of domains is $\sim \mathcal{D}^3$ and $g_1(x)$ is effectively averaged over $\sim \mathcal{D}^2$ 1D distributions.) 

\begin{figure*} [tbp]
\includegraphics[width=\textwidth]{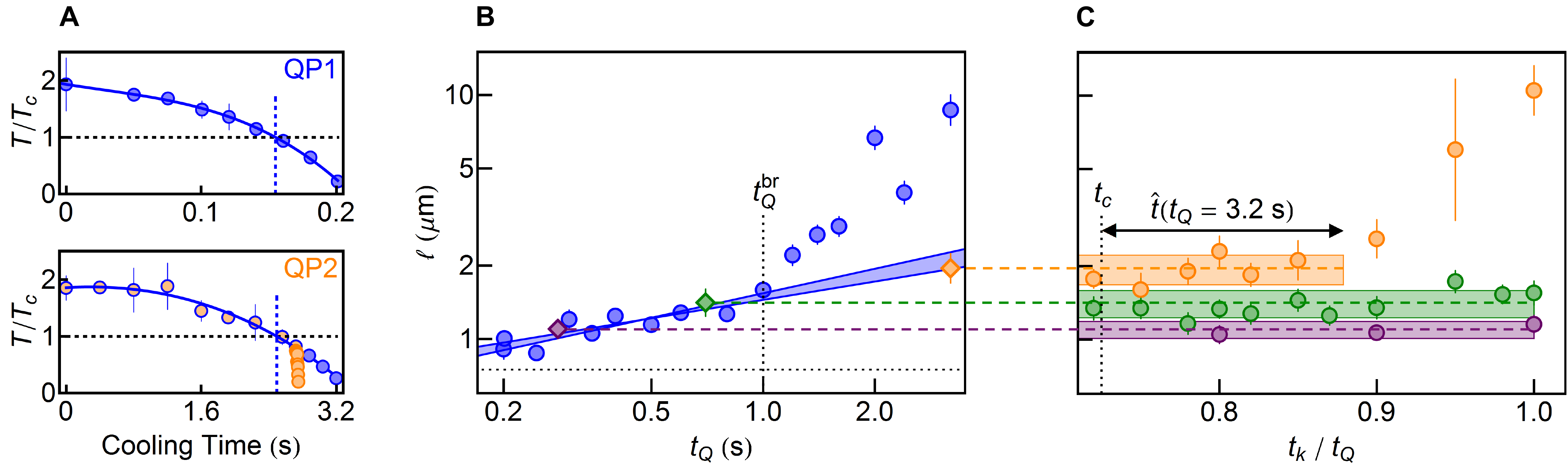}
\caption{ {\bf Kibble-Zurek scaling and freeze-out hypothesis}. (A) Quench protocols. The self-similar QP1 trajectories  are shown in blue for total cooling time $t_Q=0.2$~s (upper panel) and $3.2$~s (lower panel). 
We use polynomial fits to the data (such as shown by the solid lines) to deduce $t_c$ and $\tau_Q$.
QP2 is shown in the lower panel by the orange points, with the kink at $t_k=0.85~t_Q$. (B) Coherence length $\ell$ as a function of $t_Q$. Blue points correspond to QP1. The shaded blue area shows power-law fits with $1/4<b<1/3$ to the data with $t_Q \leq t^{\rm{br}}_Q  =1$~s. The horizontal dotted line indicates our instrumental resolution. (C) Coherence length $\ell$ measured following QP2, as a function of $t_k/t_Q$, for $t_Q= 3.2$~s (orange), $0.7$~s (green), and $0.3$~s (purple). The shaded areas correspond to the essentially constant $\ell$ (and its uncertainty) in the freeze-out period $t_k - t_c < \hat{t}$. (For $t_Q < t^{\rm{br}}_Q$ the system never unfreezes). The (average) $\ell$ values within these plateaux are shown in their respective colours as diamonds in panel B.}
\label{fig:QP}
\end{figure*}

We now turn to a quantitative study of $\ell$ for different quench protocols (Fig.~\ref{fig:QP}).
For the KZ scaling law of  Eqs.~(\ref{eq:KZ}-\ref{eq:b}) to hold, a crucial assumption is that the correlation length is essentially frozen near $t_c$. Specifically, for $\nu z=1$, which in our case holds at both MF and beyond-MF level, the freeze-out time of $\ell$ for $t>t_c$ is expected to be~\cite{Zurek:1985}
\begin{equation}
\hat{t} = f \sqrt{\tau_Q \tau_0} \, ,
\label{eq:t_hat}
\end{equation}
where $f$ is a dimensionless number of order unity. While intuitively appealing, this assumption is in principle only approximative, and the dynamics of the system coarsening (i.e., merging of the domains) at times $t>t_c$ is still a subject of theoretical work~\cite{Biroli:2010}. Practically, a crucial question is when one should measure $\ell$ in order to verify the universal KZ scaling. We resolve these issues by using two different quench protocols outlined in Fig.~\ref{fig:QP}A, which allow us both to observe the KZ scaling and to directly verify the freeze-out hypothesis, without {\it a priori} knowledge of the exact values of $f$ and $\tau_0$.

In the first quench protocol (QP1), we follow cooling trajectories such as shown in Fig.~\ref{fig:QP}A, and vary only the total cooling time $t_Q$. We restrict $t_Q$ to values between $0.2$~s and $3.5$~s, for which we observe that the cooling curves are self-similar (as seen in Fig.~\ref{fig:QP}A). We always cross $T_c =70(10)$~nK at $t_c = 0.72(5)~t_Q$ (vertical dashed line) and always have the same atom number (within $\pm 20\%$) at the end of cooling. The self-similarity of the measured cooling trajectories and the essentially constant evaporation efficiency indicate that for this range of $t_Q$ values the system is always sufficiently thermalized, the temperature (as determined from the thermal wings in ToF) is well defined during the quench~\cite{footnote:collisionless}, and to a good approximation $\tau_Q$ is simply proportional to $t_Q$. (For $t_Q <0.2$~s the evaporation is less efficient and the cooling trajectories are no longer self-similar.)

In Fig.~\ref{fig:QP}B we plot $\ell$ vs. $t_Q$, measured using QP1 (blue points). For $t_Q \leq 1$~s we observe a slow power-law growth of $\ell$, in good agreement with the expected KZ scaling (blue shaded area). However, for longer $t_Q$ this scaling breaks down and $\ell$ grows faster, quickly approaching the system size. Importantly, this breakdown can also be fully understood {\it within} the KZ framework. 
We note that the time between crossing $T_c$ and the end of cooling is $t_Q - t_c \approx 0.28\, t_Q \propto t_Q$, while the KZ freeze-out time is $\hat{t} \propto \sqrt{\tau_Q} \propto \sqrt{t_Q}$, so for slow enough quenches $t_Q - t_c$ inevitably exceeds $\hat{t}$. Hence, while it may be impossible to adiabatically cross $T_c$, in practice the system can unfreeze and heal significantly before it is observed~\cite{footnote:harmonic_b}. From the point where the KZ scaling breaks  down in Fig.~\ref{fig:QP}B, $t^{\rm{br}}_Q\approx1$~s, we posit that for $t_Q=t^{\rm{br}}_Q$ we have $\hat{t} \approx 0.28 \, t^{\rm{br}}_Q$ and hence, from Eq.~(\ref{eq:t_hat}), more generally $\hat{t} \approx 0.28 \sqrt{t_Q t^{\rm{br}}_Q}$.

To verify this picture, we employ a second quench protocol (QP2), which involves two cooling steps, as shown by the orange points in the bottom panel of Fig.~\ref{fig:QP}A. We initially follow the QP1 trajectory for a given $t_Q$, but then at a variable ``kink" time $t_k \gtrsim t_c$ we accelerate the cooling; the last part of the trajectory always corresponds to the final portion of our fastest, $0.2$-s cooling trajectory.  This way, even for $t_Q > t^{\rm{br}}_Q$ we can complete the cooling and measure $g_1$ before the system has time to unfreeze.

In Fig.~\ref{fig:QP}C, the orange points show the QP2 measurements of $\ell$ for $t_Q=3.2$~s and various values of the kink position $t_k/t_Q$. These data reveal two remarkable facts. First, for a broad range of $t_k$ values, $\ell$ is indeed constant (within errors), and the width of this plateau agrees with our estimate $\hat{t} \approx 0.5$~s for $t_Q=3.2$~s, indicated by the horizontal arrow.  Second, the value of $\ell$ within the plateaued region falls in line with the KZ scaling law in Fig.~\ref{fig:QP}B.
We also show analogous QP2 measurements for $t_Q = 0.7$~s (green) and $0.3$~s (purple); in these cases $t_Q < t^{\rm{br}}_Q$, so $\hat{t}$ is longer than $t_Q - t_c $, the system never unfreezes, and thus the acceleration of the cooling has no effect on $\ell$. These results provide direct support for the KZ freeze-out hypothesis.

\begin{figure} [b]
\centerline{\includegraphics[width=0.5\columnwidth]{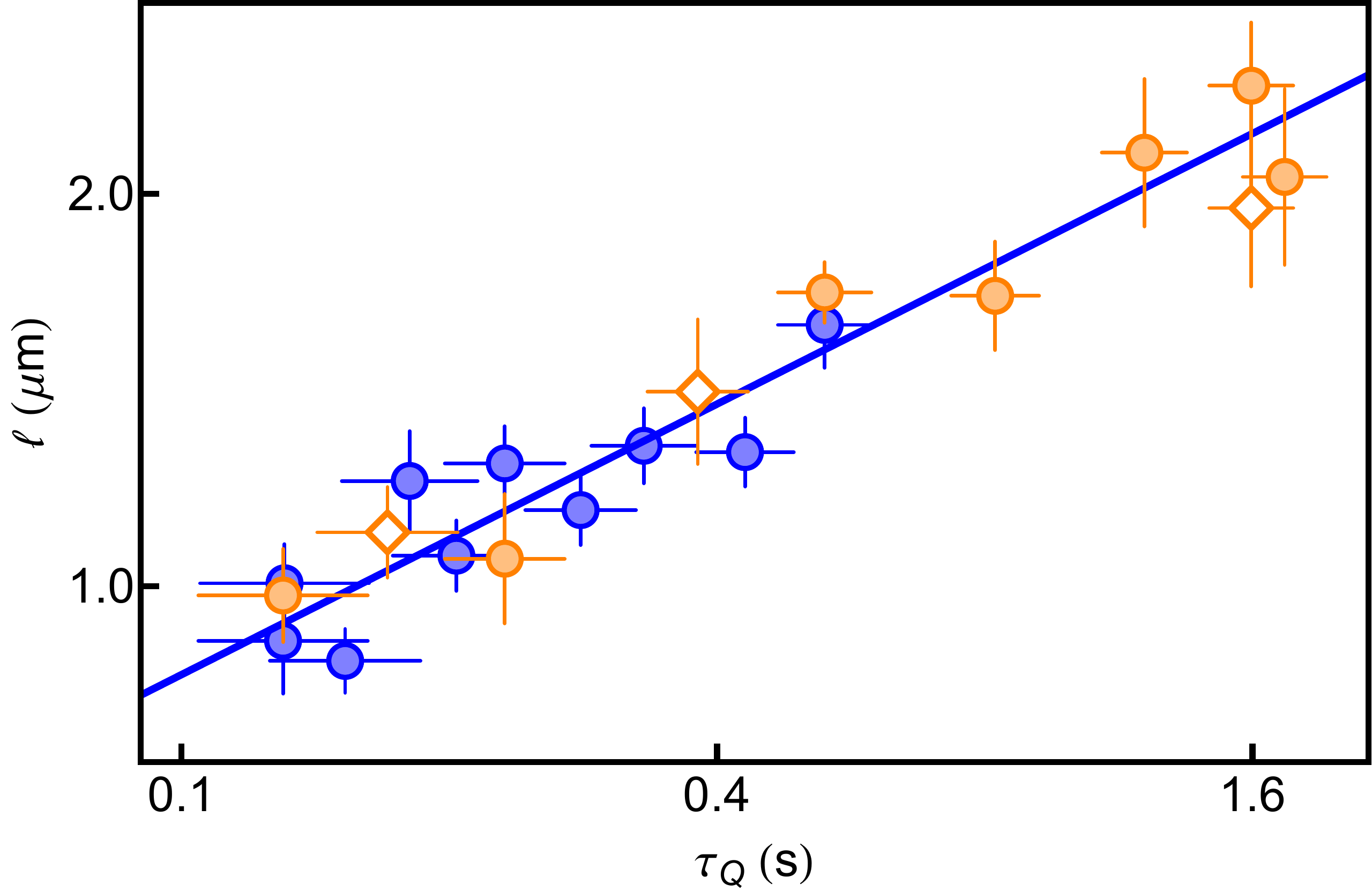}}
\caption{ {\bf Critical exponents of the interacting BEC transition}. Orange circles and diamonds show $\ell$ values obtained using QP2,  as in Fig.~\ref{fig:QP}C; the diamonds show the same three data points as in Fig.~\ref{fig:QP}B.  Blue circles show the same QP1 data with $t_Q\leq1$~s as in  Fig.~\ref{fig:QP}B. We obtain $b=0.35(4)$ (solid line), in agreement with the F-model prediction $b\approx1/3$, corresponding to $\nu\approx 2/3$ and $z=3/2$, and excluding the mean-field value $b=1/4$. }
\label{fig:Master}
\end{figure}

To accurately determine the KZ exponent $b$, we have made extensive measurements following QP2, extracting $\ell$ from the plateaued regions of width $\min[\hat{t}, t_Q-t_c]$, as in Fig.~\ref{fig:QP}C. In Fig.~\ref{fig:Master} we combine these data with the QP1 measurements for $t_Q \leq 1$~s, and plot  $\ell$ versus $\tau_Q$. The plotted values of $\tau_Q$ and their uncertainties include the small systematic variation of the derivative of our cooling trajectory between $t_c$ and $t_c - \hat{t}$.  We finally obtain $b=0.35(4)$, which strongly favors the F-model prediction $b\approx1/3$ over the MF value $b=1/4$~\cite{footnote:rescaling}.

Having observed excellent agreement with the KZ theory, we discuss the implications of our measurements for the critical exponents of the interacting BEC phase transition, which is in the same universality class as the $\lambda$-transition of $^4$He. 
While $\nu\approx 0.67$ has been measured in both liquid helium~(see~\cite{Pogorelov:2007}) and atomic gases~\cite{Donner:2007}, the dynamical exponent $z$ has, to our knowledge, never been measured before (see \cite{Hohenberg:1977,Folk:2006}).
Using the well-established $\nu=0.67$ and  Eq.~(\ref{eq:b}), we obtain $z=1.4(4)$. In contrast, MF theory does not provide a self-consistent interpretation of our results, since fixing $\nu=1/2$ yields an inconsistent $z=0.9(4)$.
Interestingly, if we instead fix $\nu z=1$, which holds at both MF and F-model level, from Eq.~(\ref{eq:b}) we obtain a slightly more precise $z=1.4(2)$ and also recover $\nu = 0.70(8)$.

In the future, it would be interesting to study the effect of tuneable interactions in a homogeneous atomic gas on the value of $b$. According to the Ginzburg criterion, near the critical point MF breaks down for $\xi \gtrsim \xi_{\rm G} = \lambda_c^2/(\sqrt{128}\pi^2 a)$, where $a$ is the s-wave scattering length. It is tempting to combine the (dynamical) KZ and (equilibrium) Ginzburg arguments and speculate that one should observe the MF value of $b$ if on approach to $T_c$ the KZ freeze-out occurs before MF breaks down, and the F-value of $b$ if the reverse is true. In our experiments $\xi_{\rm G} \approx 0.8~\mu$m and the freeze-out values of $\ell$ are systematically higher. However, with use of a Feshbach resonance the opposite regime should be within reach. Another interesting future study could focus on the dynamics of domain coarsening.
Finally, our methods could potentially be extended to studies of higher-order correlation functions and the full statistics of the domain sizes.



\begin{scilastnote}
\item We thank Martin Robert-de-Saint-Vincent for experimental assistance, Richard Fletcher for comments on the manuscript, and Nigel Cooper, Jean Dalibard, Gabriele Ferrari, Bill Phillips, and Wilhelm Zwerger for insightful discussions.
This work was supported by AFOSR, ARO, DARPA OLE, and EPSRC [Grant No. EP/K003615/1]. N.N. acknowledges support from Trinity College, Cambridge, and R.P.S. from the Royal Society. 

\end{scilastnote}


\newpage

\newpage

\newpage

\cleardoublepage

\baselineskip24pt

\bigskip
\noindent{Supplementary Material for}\\
\noindent{\textbf{\large{Critical Dynamics of Spontaneous Symmetry Breaking in a Homogeneous Bose gas}}\\
\noindent{Nir Navon$^{\ast}$, Alexander L. Gaunt$^{\ast}$, Robert P. Smith, Zoran Hadzibabic}\vspace{0.3cm}

\smallskip
\noindent\textbf{Measurement of the first-order correlation function $g_1(x)$}

Here we detail the method to measure the first-order correlation function $g_1(x)$ using the interferometric scheme of Fig. 2A (main text). 
The principle of the measurement is to use a displaced copy of the cloud as a homodyne phase reference and probe via interference the phase coherence of the gas at rest. It involves four steps: (i) a coherent superposition of the cloud at rest and its copy moving along $\hat{\bf x}$ is created by a resonant Bragg pulse, (ii) the moving copy is allowed to shift by a variable distance, (iii) the interference is achieved by a second Bragg pulse, and (iv) the Bragg-diffracted atoms are allowed to separate from the main cloud during a long time-of-flight (ToF). 

Let us consider the gas initially at rest. A Bragg pulse couples atoms of momentum ${\bf q}$ to a state of momentum ${\bf q}+{\bf q}_r$, where ${\bf q}_r$ is the recoil momentum. For our short pulses the efficiency of this coupling is independent of ${\bf q}$. Moreover, in our case the spread of $|{\bf q}|$ is $\ll |{\bf q}_r|$, so after a long flight time we can fully distinguish the atoms at rest and the recoiling ones. We can thus reduce the problem to a fictitious two-level system, where the effective state $\ket{0}$ corresponds to the atoms (approximately) at rest, and $\ket{{\bf q}_r}$ to the atoms moving at (approximately) recoil velocity. The fact that the recoil velocity, ${\bf v}_r = \hbar {\bf q}_r/m $,  is much larger than any other characteristic velocity in the system also means that to a good accuracy we can assume that the intrinsic evolution of the system between the two Bragg pulses is negligible.

In our basis, the initial state of the cloud is 
\be
\ket{\Psi_0}=\begin{pmatrix}\psi({\bf r})\\0\end{pmatrix},
\ee
where $\psi({\bf r})$ is the wavefuction describing the cloud at rest (which vanishes outside the box trap) and $\int {\rm d}{\bf r}|\psi|^2=N$. 

(i) The Bragg pulse of area $\theta$ corresponds to the operator 
\be
\hat{B}_\theta=\begin{pmatrix}\cos\left(\theta/2\right) & i\sin\left(\theta/2\right) \\
i\sin\left(\theta/2\right) & \cos\left(\theta/2\right) 
\end{pmatrix} \, ,
\ee
so after the first pulse the state of the system is
\be
\ket{\Psi'}=\begin{pmatrix}\psi({\bf r})\cos(\theta/2)\\i\psi({\bf r})\sin(\theta/2)\end{pmatrix}.
\ee

(ii) In between the two Bragg pulses, the recoiling copy of the cloud is displaced with respect to the stationary one. This is described by the free evolution operator 
\be
\hat{U}_{\Delta t}=\begin{pmatrix}1 & 0 \\
0 & \hat{T}_{\Delta t}
\end{pmatrix},
\ee
where $\hat{T}_{\Delta t}\phi({\bf r})=\phi({\bf r}-{\bf v}_r \Delta t)$ is the translation operator. Hence, the state of the system just before the second pulse is
\be
\ket{\Psi''}=\begin{pmatrix}\psi({\bf r})\cos(\theta/2)\\i\psi({\bf r}-{\bf v}_r \Delta t)\sin(\theta/2)\end{pmatrix}.
\ee

(iii) Applying the Bragg operator again, the state of the system after the second pulse is
\be \label{Psifinal}
\ket{\Psi}=\hat{B}_\theta \hat{U}_{\Delta t} \hat{B}_\theta \ket{\Psi_0}=\begin{pmatrix} \psi({\bf r})\cos^2(\theta/2)-\psi({\bf r}-{\bf v}_r \Delta t)\sin^2(\theta/2) \\ [\psi({\bf r})+\psi({\bf r}-{\bf v}_r \Delta t)]i\sin(\theta/2)\cos(\theta/2) \end{pmatrix}.
\ee
Thus, the expectation value for the final number of recoiling atoms (obtained by averaging over many realizations) is 
\be \label{Nrexp}
N_r = \frac{\sin^2\theta}{4}\int\; {\rm d}{\bf r} \left\langle |\psi({\bf r}) + \psi({\bf r}-{\bf v}_r \Delta t)|^2 \right\rangle \, .
\ee
We now write $\left \langle |\psi({\bf r})+\psi({\bf r}-{\bf v}_r \Delta t)|^2 \right \rangle = |\psi({\bf r})|^2 + |\psi({\bf r}-{\bf v}_r \Delta t)|^2 + 2 \operatorname{Re}\left[ \left\langle \psi^*({\bf r}) \psi({\bf r}-{\bf v}_r \Delta t) \right\rangle \right]$. For each of the first two terms the integral in Eq.~(\ref{Nrexp}) simply gives $N$. For the third, interference term, we note that:

\textbullet \; if both ${\bf r}$ and ${\bf r}-{\bf v}_r \Delta t$ are inside the box, $\operatorname{Re}\left[ \left\langle \psi^*({\bf r}) \psi({\bf r}-{\bf v}_r \Delta t) \right\rangle \right] = n g_1(x)$, where $n$ is the uniform density of the trapped gas and $x=|{\bf v}_r|\Delta t$, and

\textbullet \; $\operatorname{Re}\left[ \left\langle \psi^*({\bf r}) \psi({\bf r}-{\bf v}_r \Delta t) \right\rangle \right] = 0$ otherwise. \\
Considering these two cases, we evaluate the integral in Eq.~(\ref{Nrexp}) to recover the result of Eq.~(4) in the main text:
\begin{equation}
\frac{N_r}{N} = \frac{1}{2} \left[ 1+ \left(  1- \frac{x}{L}\right) g_1(x) \right] \sin^2 \theta \,.
\end{equation}

(iv) Experimentally, we measure $N_r/N$ by allowing the recoiling atoms to fully separate from the stationary cloud in a long (140~ms) ToF, as illustrated in Fig.~S1. We repeat each measurement several times and to get $g_1(x)$ also vary the time separation between the Bragg pulses, $\Delta t$. 

Interestingly, in our system it is not possible to directly optically resolve the KZ domains (as sketched {\it in situ} in Fig. 2A in the main text), but we can still deduce their size by simply counting the recoiling atoms as explained here.

\begin{figure}[h!]
\centerline{\includegraphics[width=\columnwidth]{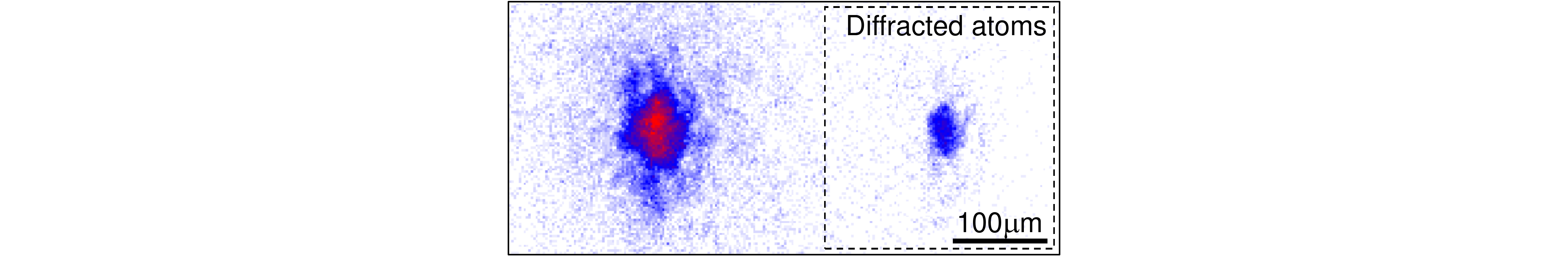}}
\caption*{Fig. S1. Example of an experimental image used for the measurement of $g_1(x)$. For this particular image, the quench time was $t_Q=0.8$~s and the separation between the two Bragg pulses was $\Delta t=0.4$~ms. A typical measurement of a single $g_1(x)$ function involves 4 repetitions of the experiment at 20 different Bragg-pulse separation times. }\label{FigS1}
\end{figure}


\begin{thebibliography}{10}
\bibitem{footnote:collisionless}
Near $T_c$ the mean free path for classical elastic collisions is about four times larger than the size of the box. Hence, while the evaporation takes place at the box walls, we can also safely assume that $T$ is uniform across the sample.
\bibitem{footnote:rescaling}
Note that scaling $\ell$ to $\lambda_c$ and $\tau_Q$ to $\tau_{\rm el}$, for each data series separately, affects $b$ by $<0.01$.
\end{thebibliography}
\end{document}